\documentclass[prd,twocolumn,amsmath,amssymb]{revtex4}
\usepackage{graphicx}

\def\al{\alpha}
\def\be{\beta}
\def\ga{\gamma}
\def\de{\delta}
\def\ep{\epsilon}
\def\ve{\varepsilon}

\def\th{\theta}

\def\ka{\kappa}
\def\la{\lambda}

\def\si{\sigma}

\def\ph{\phi}

\def\om{\omega}

\def\Om{\Omega}

\def\prt{\partial}
\def\cl{{\cal L}}

\def\Frac#1#2{{\textstyle{{#1}\over {#2}}}}
\def\half{{\textstyle{1\over 2}}}
\def\lsim{\mathrel{\rlap{\lower4pt\hbox{\hskip1pt$\sim$}}
    \raise1pt\hbox{$<$}}}
\def\gsim{\mathrel{\rlap{\lower4pt\hbox{\hskip1pt$\sim$}}
    \raise1pt\hbox{$>$}}}
\def\sqr#1#2{{\vcenter{\vbox{\hrule height.#2pt
         \hbox{\vrule width.#2pt height#1pt \kern#1pt
         \vrule width.#2pt}
         \hrule height.#2pt}}}}
\newcommand{\beq}{\begin{equation}}
\newcommand{\eeq}{\end{equation}}
\newcommand{\bea}{\begin{eqnarray}}
\newcommand{\eea}{\end{eqnarray}}
\newcommand{\rf}[1]{(\ref{#1})}

\def\mbf#1{\mbox{\boldmath$#1$}}
\def\syjm#1#2{\phantom{}_{#1}Y_{#2}}

\def\kfd#1{k_{F}^{(#1)}}
\def\kafd#1{k_{AF}^{(#1)}}
\def\kjm#1#2#3{k^{(#1)}_{(#2)#3}}

\def\kVdjm#1#2{\kjm{#1}{V}{#2}}
\def\k{\kafd{3}}
\def\cg#1#2{\langle #1 | #2 \rangle}
\def\EzE{E^{(0E)}_{jm}}
\def\EoE{E^{(1E)}_{jm}}
\def\EoB{E^{(1B)}_{jm}}
\def\BzB{B^{(0B)}_{jm}}
\def\BoB{B^{(1B)}_{jm}}
\def\BoE{B^{(1E)}_{jm}}
\def\EzEjm#1{E^{(0E)}_{#1}}
\def\EoEjm#1{E^{(1E)}_{#1}}
\def\EoBjm#1{E^{(1B)}_{#1}}

\def\kzE{{\mathcal K}^{(0E)}_{jm}}
\def\koE{{\mathcal K}^{(1E)}_{jm}}
\def\koB{{\mathcal K}^{(1B)}_{jm}}
\def\kzEjm#1{{\mathcal K}^{(0E)}_{#1}}

\def\dr{\Frac{\prt\phantom{r}}{\prt r}}
\def\ddr{\Frac{\prt^2\phantom{r}}{\prt r^2}}

\begin{document}

\title{Bounds on Lorentz and CPT Violation from the Earth-Ionosphere Cavity}
\author{Matthew Mewes}
\affiliation{Physics Department, Marquette University,
  Milwaukee, WI 53201, U.S.A.}
\date{\today}

\begin{abstract}
  Electromagnetic resonant cavities 
  form the basis of many tests
  of Lorentz invariance involving photons. 
  The effects of some forms of
  Lorentz violation
  scale with cavity size.
  We investigate possible
  signals of violations
  in the naturally occurring
  resonances formed in the
  Earth-ionosphere cavity.
  Comparison with observed resonances
  places the first terrestrial constraints
  on coefficients associated with
  dimension-three Lorentz-violating operators
  at the level of $10^{-20}$ GeV.
\end{abstract}
\maketitle

\section{Introduction}

Modern versions
\cite{cavities}
of the classic
Michelson-Morley 
\cite{mm}
and Kennedy-Thorndike
\cite{kt}
experiments are among the most
sensitive tests of Lorentz invariance,
the symmetry behind special relativity.
Typically
these experiments search for minute
changes in the resonant frequencies of
electromagnetic cavities.
High quality factors allow for precise
tracking of the frequencies,
giving extreme sensitivities to
possible deviations from perfect
Lorentz symmetry.
However,
the effects of some forms of Lorentz violation
increase with wavelength.
As a result,
high sensitivities may be achieved
using very-low-frequency resonances,
such as those that naturally occur
in the Earth-ionosphere cavity,
despite their relatively low
quality factors.

In this work,
we consider signals of Lorentz violation
that would appear in Schumann resonances,
the lowest-frequency standing waves that
form in the atmosphere
\cite{schum1,schum2}.
We obtain conservative bounds by
comparison with observations
\cite{schumdata}.
The Earth's surface and ionosphere
form a cavity of immense size,
leading to resonances with 
very long wavelengths.
The lowest-frequency resonances
have wavelengths that are comparable to
the circumference of the Earth
and have frequencies as low as 8 Hz.

Violations of Lorentz invariance
are described by the
Standard-Model Extension (SME)
\cite{ck,kost}.
The SME is a theoretical framework
that provides a basis
for many experimental and
theoretical studies of Lorentz violation
\cite{cpt,smetables},
including those involving
atoms \cite{ccexpt,spaceexpt},
hadrons \cite{hadrons,hadronth},
fermions \cite{electrons,electronth,muons,neutrinos},
the Higgs boson \cite{higgs},
gravity \cite{gravity},
and photons
\cite{cavities,photons,km_cmb,km_astro,bire,km}.
In addition to resonant-cavity experiments,
searches for Lorentz violation in photons
using the SME approach include
astrophysical searches
for vacuum birefringence
\cite{bire,km_cmb,km_astro,km}
and dispersion 
\cite{km_astro,km}.
The goal of the SME is the characterization
of all violations of Lorentz symmetry that
are consistent with known physics
using effective field theory
\cite{kpo}.
While motivated in part
by the possibility of spontaneous
symmetry breaking in strings
\cite{ks,kp},
it encompassed violations
with other origins
\cite{ncqed,coup,qg,fn,bj,ss,bluhm,gm}.
Much of the work on Lorentz violation
has focused on the minimal SME,
which includes operators of renormalizable
dimension in a flat spacetime.
While nonrenormalizable operators
and curved spacetimes are of general interest
\cite{kost,coup,gravity,km_cmb,km_astro,kle},
Schumann resonances are particularly 
sensitive to the dimension-three $CPT$-odd
Lorentz-violating operators of the 
minimal-SME photon sector.

In cavities,
Lorentz violation can introduce
frequencies that depend on the
orientation of the cavity,
signalling rotation violations,
and dependence on velocity resulting
from boost violations
\cite{km}.
The quantity determining their sensitivity
is the dimensionless
fractional frequency shift $\de\nu/\nu$.
To date, cavity experiments have
focused primarily on one class of violations,
namely the dimension-four $CPT$-even
Lorentz-violating operators
of the minimal SME.
The coefficients associated with
these operators $(\kfd{4})^{\al\be\ga\de}$
are dimensionless.
Therefore,
dimensional analysis
suggests frequency shifts of the form
$\de\nu/\nu \sim (\kfd{4})^{\al\be\ga\de}$,
which implies little or no dependence on frequency.
Consequently,
there is little advantage to using
low-frequency resonances.
In contrast,
the coefficients associated with
the dimension-three operators
$(\k)_\ka$ have mass-dimension one.
Therefore, we naively expect
shifts in frequency that depend on the
ratio $(\k)_\ka/\nu$.
Given that $\nu\sim 10^{-23}$ GeV
for Schumann resonances,
we naively expect sensitivities
on the order of $10^{-23}$ GeV to $\k$
coefficients,
assuming at least order-one
sensitivity to $\de\nu/\nu$.
While not as sensitive as
birefringence tests
\cite{km_astro},
the bounds obtained here
represent the first terrestrial
bounds on the dimension-three operators,
providing a valuable check
on existing astrophysical constraints.

The structural outline of this paper
is as follows.
Section \ref{sec_bg} provides some
basic theory behind our calculation.
In Sec.\ \ref{sec_cal} we derive
modified wave equations and describe
a numerical method of determining
the effects of Lorentz violations
on Schumann resonances.
The results of our calculation
are discussed in Sec.\ \ref{sec_results}.
Unless otherwise stated,
we use the notation and conventions
of Refs.\ \cite{ck,km_astro}.

\section{Background}
\label{sec_bg}

In this section we discuss
the theory behind the calculation
of the Earth-ionosphere resonances 
in the presence of dimension-three
Lorentz violations.
We begin by discussing the
conductivity of the atmosphere
and the Schumann resonances
in the usual case.
We then review the modified
electrodynamics including
the $CPT$-odd operators of
the minimal SME.

\subsection{Conductivity profile}

Resonances in the Earth-ionosphere cavity
are excited by a number of man-made
and natural phenomena,
lightning being a primary source.
The surface of the Earth forms the
lower boundary and,
in our calculation,
is treated as a perfect conductor.
The ionosphere forms a lossy upper boundary
with a finite conductivity profile
that increases with altitude.
The conductivity of the lower atmosphere
can be approximated by a ``knee''-model
that separates into two layers
with exponentially increasing conductivity
\cite{knee}:
\begin{align}
  \si(r) &\simeq
  \left\{\begin{array}{ll}
  \infty \quad & r<R \ , \\
  \si_0 \exp{\Frac{r-r_0}{\xi_l}} \quad & R<r<r_0 \ , \\
  \si_0 \exp{\Frac{r-r_0}{\xi_u}} \quad & r_0<r \ ,
  \end{array}\right.
\end{align}
where
$r$ is the distance for the center of the Earth,
and 
$R \simeq 6400\ {\rm km }
\simeq 3.2 \times 10^{22}\ {\rm GeV}^{-1}$
is the Earth radius.
The lower layer is dominated by positive
and negative ions.
The upper layer approximates
the bottom of the ionosphere,
which is dominated by free electrons.
The knee radius $r_0$ is the boundary
between the layers,
and $\si_0$ is the conductivity
at this transition.
In units where $c=\hbar=\ep_0=\mu_0=1$,
we adopt values
$r_0=1.009 R$,
$\xi_l=0.007 R$,
$\xi_u=0.0005 R$, and
$\si_0=1.3 R^{-1}$
in the numerical calculations that follow.
This profile yields frequencies
and quality factors that closely
match the observed resonances in the
Lorentz-invariant limit.

In the conventional case,
both transverse-magnetic (TM) and
transverse-electric (TE) modes 
may be excited in the cavity.
However,
TE modes oscillate in the radial direction,
implying wavelengths comparable to the
height of the ionosphere,
yielding frequencies in the kHz range.
In contrast,
the lowest-frequency TM modes
vary little in the radial direction
but form standing waves that encircle the Earth.
As a result,
the wavelengths are comparable
to the Earth's circumference,
yielding frequencies as low as 8 Hz.
Our goal is to understand the effects
of Lorentz violations on these
low-frequency Schumann resonances.

\subsection{Modified electrodynamics}

The lagrangian governing
electromagnetic waves,
including dimension-three
Lorentz-violating operators,
is given by 
\cite{ck}
\beq
\cl = -\Frac14 F_{\mu\nu}F^{\mu\nu}
+\half\ep^{\ka\la\mu\nu}(\k)_\ka A_\la F_{\mu\nu} \ ,
\label{lagrangian}
\eeq
where $F_{\mu\nu}=\prt_\mu A_\nu - \prt_\nu A_\mu$
is the field strength.
The resulting theory preserves
the usual gauge symmetry,
but violates $CPT$ invariance.
Lorentz and $CPT$ violations are
controlled by the constant
coefficients $(\k)_\ka$,
which include a pseudoscalar $(\k)_0$
and pseudovector $\mbf\k$.
The $(\k)_\ka$ coefficients
are assumed to be constant,
which leads to energy-momentum
conservation.
More generally one can consider
Lorentz-violating backgrounds
with spacetime variations.
These types of violations
are particularly important
when considering Lorentz
violations in curved spacetimes
\cite{kost,gravity}.
However,
in scenarios where violations
originate in the moments
shortly after the big bang,
it is likely that any variation
has expanded,
leading to little fluctuation
over experimentally relevant
time and length scales.
Consequently,
the idea presented here probes $(\k)_\ka$
in our local neighborhood.
Astrophysical tests rely on light
that has propagated across much of
the visible universe.
So they test Lorentz
invariance over much larger scales
and could be drastically affected
by variations in $(\k)_\ka$.
Therefore,
while current bounds from
astrophysical searches for birefringence
currently lie at the $10^{-42}$ GeV level
\cite{km_astro},
terrestrial tests provide an important
complementary set of constraints.

The equations of motion resulting
from Eq.\ \rf{lagrangian} provide
Lorentz-violating inhomogeneous
Maxwell equations.
We are primarily interested
in harmonic solutions
and consider electric and magnetic
fields of the form
$\mbf E(t) =\mbf E(\om) e^{-i\om t},
\mbf B(t) =\mbf B(\om) e^{-i\om t}$.
Together with the usual
homogeneous equations,
we arrive at a Lorentz-violating
electrodynamics with a
modified Amp\'ere law and a 
conventional Faraday law: 
\begin{align}
  \hspace*{-3pt}&
  \mbf\nabla\times\mbf B +i\om\mbf E - 2\, (\k)_0\, \mbf B + 2\, \mbf\k\times\mbf E
  -\si\mbf E = \mbf J_s \ , \notag \\
  \hspace*{-3pt}&
  \mbf\nabla\times\mbf E -i\om\mbf B = 0 \ .
  \label{max1}
\end{align}
Here we have included the usual
source current and conductivity terms.
These conventional source terms result
if we assume usual coupling to matter.
The source current is not needed
in determining the resonances
but could be used in modeling the effects
of individual lightning strikes.
The conductivity term is necessary to obtain
realistic frequencies and quality factors,
which are affected by the
profile of the upper boundary
and the losses it introduces.

\section{Calculation}
\label{sec_cal}

Our goal is to
calculate the resonant frequencies 
that result from the modified Maxwell equations.
We begin by expanding into spherical
harmonics and deriving modified wave equations.
A numerical calculation is used to
estimate the resulting resonant frequencies.
The results are discussed
in the following section.

\subsection{Wave equations}

We begin our search for resonances
by first expressing the modified Maxwell
equations \rf{max1} 
in spherical coordinates using
the helicity-basis and identities
discussed in Appendix \ref{sec_sw}.
This involves writing the
Maxwell equations in covariant form,
then using relation \rf{J-ident2}
to express them in terms of
$\prt/\prt r$
and covariant angular-momentum
ladder operators $J_\pm$.
Dropping the source current,
the result is
\begin{align}
  0 &= J_+B_-+J_-B_+ +(\om+i\si)r E_r
  +2ir(\k)_0B_r 
  \notag \\ &\quad 
  + 2r(\k)_+E_- - 2r(\k)_-E_+  \ , 
  \label{amp_r} \\
  0 &= \pm \dr r B_\pm
  - J_\pm B_r +(\om+i\si)r E_\pm
  +2ir(\k)_0 B_\pm 
  \notag \\ &\quad 
  \pm 2r(\k)_r E_\pm \mp 2r(\k)_\pm E_r  \ , 
  \label{amp_pm} \\
  0 &= J_+ E_- +J_- E_+ -\om r B_r  \ , 
  \label{far_r} \\
  0 &= \pm \dr r E_\pm 
  - J_\pm E_r -\om r B_\pm \ , 
  \label{far_pm}
\end{align}
where $E_r$, $B_r$ are
the radial field components,
and $E_\pm$, $B_\pm$ are
negative/positive helicity components,
as discussed in the appendix.
The components of $\mbf\k$ are defined by
$(\k)_a=\mbf{\hat e}_a\cdot \mbf\k$,
where ${\hat e}_a$ are the helicity
basis vectors.
The components $E_\pm$ and $B_\pm$ are
referred to as spin-weighted functions
with spin-weight-$(\pm 1)$,
while $E_r$, $B_r$ have a weight of zero.
Spin weight and helicity are 
equivalent up to a sign.
Eqs.\ \rf{amp_r} and \rf{far_r} provide
two scalar relations, while
\rf{amp_pm} and \rf{far_pm}
have a spin weight of $\pm 1$.

We can expand the field components in
spin-weighted spherical harmonics.
These provide the generalization of
the familiar spherical harmonics
to spin-weighted functions.
The expansion takes the form
\begin{align}
  E_r &= \sum \Frac{1}{r} \EzE \, \syjm{0}{jm} \ ,
  \\
  E_\pm & = \sum \sqrt{\Frac{j(j+1)}{2}}\Frac{1}{r}
  (\mp \EoE - i \EoB ) \, \syjm{\pm1}{jm} \ , 
  \\
  B_r &= \sum \Frac{1}{r} \BzB \, \syjm{0}{jm} \ ,
   \\
  B_\pm & = \sum \sqrt{\Frac{j(j+1)}{2}}\Frac{1}{r}
  (\mp \BoB - i \BoE ) \, \syjm{\pm1}{jm} \ ,
\end{align}
where
$\EzE$, $\EoE$, $\EoB$,
$\BzB$, $\BoB$, and $\BoE$
are $r$-dependent field coefficients.
They are associated with
total-angular-momentum eigenmodes
and have $E$-type parity, $(-1)^j$, or
$B$-type parity, $(-1)^{j+1}$.
The $\sqrt{j(j+1)/2}$ and $1/r$
factors in the expansions are
for convenience.

Using these expansions,
we can express the Maxwell equations
in terms of 
$\EzE$, $\EoE$, $\EoB$,
$\BzB$, $\BoB$, and $\BoE$.
First, using the ladder operators $J_\mp$
to lower/raise the spin-$(\pm 1)$ relations
\rf{amp_pm} and \rf{far_pm}
we arrive at six scalar Maxwell equations.
We then use the spherical-harmonic
expansions of the fields and the 
orthogonality relation \rf{orth}
to get relations between 
the six expansion coefficients.
Some algebra yields
three $E$-parity equations
and three $B$-parity equations:
\begin{align}
  0 &= j(j+1)\BoE - i(\om+i\si)r\EzE + r \kzE \ , \label{E2} \\
  0 &= \dr \BoE - i(\om+i\si) \EoE + \Frac{1}{j(j+1)}\koE \ , \label{E3} \\
  0 &= r\dr \BoB - \BzB + i (\om+i\si)r\EoB \notag \\ & \quad + \Frac{r}{j(j+1)}\koB \ , \label{B3} \\
  0 &= j(j+1) \EoB + i \om r \BzB \ , \label{B1} \\
  0 &= \dr \EoB + i\om \BoB \ , \label{B2} \\
  0 &= r \dr \EoE - \EzE  - i\om r \BoE \ , \label{E1} 
\end{align}
where the Lorentz- and $CPT$-violating 
contributions have been collected into the
field combinations
\begin{align}
  \kzE &= 2(\k)_0 \BzB + 2i|\mbf\k|m\EoE
  \notag \\ & \quad
  + 2|\mbf\k|(j-1){\mathcal C}_{jm} \EoBjm{(j-1)m}
  \notag \\ & \quad
  - 2|\mbf\k|(j+2){\mathcal C}_{(j+1)m} \EoBjm{(j+1)m} \ , 
  \label{kr} \\
  \koE &= 2(\k)_0 j(j+1)\BoB 
  \notag \\ & \quad
  +2i|\mbf\k|m\EoE
  +2i|\mbf\k|m\EzE
  \notag \\ & \quad
  +2|\mbf\k|(j^2+2j-1){\mathcal C}_{jm}\EoBjm{(j-1)m}
  \notag \\ & \quad
  +2|\mbf\k|(j^2-2){\mathcal C}_{(j+1)m}\EoBjm{(j+1)m} \ ,
  \label{kE} \\
  \koB &= -2(\k)_0 j(j+1)\BoE -2i|\mbf\k|m\EoB
  \notag \\ & \quad
  -2|\mbf\k|(j+1){\mathcal C}_{jm}\EzEjm{(j-1)m}
  \notag \\ & \quad
  +2|\mbf\k|j{\mathcal C}_{(j+1)m}\EzEjm{(j+1)m}
  \notag \\ & \quad
  +2|\mbf\k|(j^2+2j-1){\mathcal C}_{jm}\EoEjm{(j-1)m}
  \notag \\ & \quad
  +2|\mbf\k|(j^2-2){\mathcal C}_{(j+1)m}\EoEjm{(j+1)m} \ ,
  \label{kB}
\end{align}
where
${\mathcal C}_{jm} = \sqrt{(j^2-m^2)/(4j^2-1)}$.
Here we take the 
angular-momentum quantization axis along
the direction of $\mbf\k$.
Note that Eqs.\ \rf{E2}-\rf{B3} 
correspond to the modified spherical Amp\'ere law
and Eqs.\ \rf{B1}-\rf{E1} are the
usual Faraday law.

In the conventional case,
where all coefficients for Lorentz violation are zero,
rotational symmetry implies that
resonances are eigenmodes of angular momentum
with definite $j$ and $m$ values.
The symmetry also implies degeneracy in $m$.
So the different resonant frequencies
correspond to different values of
the total-angular-momentum index $j$.
Setting $\kzE=\koE=\koB=0$ in Eqs.\ \rf{E2}-\rf{E1},
we also note that the Lorentz-invariant case
splits according to parity.
The $B$-parity resonances
correspond to the high-frequency TE modes,
while low-frequency TM Schumann resonances
are the $E$-parity modes.

Allowing for Lorentz violations,
the new symmetries of the system lead
to several generic predictions.
Rotational symmetry is preserved
in the event that we have only
isotropic violations associated with
coefficient $(\k)_0$.
This implies that  the indexing and
degeneracies of the modes is the same,
but the frequencies may change.
In contrast,
the vector $\mbf\k$ breaks
the usual degeneracy.
The system remains symmetric under
rotations about $\mbf\k$.
So we expect
resonances that are
eigenmodes of these rotations
with eigenvalues $m$, as usual.
However,
these coefficients break
spherical symmetry,
implying the index $j$ is no longer
associated with resonances.
As a result,
the usual $2j+1$ degeneracies should break,
yielding modes with definite $m$
but indefinite $j$.
Consequently,
we expect two types of
effects that would signify possible
Lorentz violation.
One is a split of degeneracies leading to
additional resonant frequencies.
This results from anisotropic violations.
The other effect is a shift in frequencies
that may result from either
anisotropic and isotropic violations.

The above first-order differential
equations can be reduced to second-order
modified wave equations.
We begin by using the Faraday law,
Eqs.\ \rf{B1}-\rf{E1},
to eliminate the magnetic field.
We also use Eq.\ \rf{E2}
to eliminate the electric field
component $\EzE$ in favor of 
$\EoE$, $\EoB$, and $\kzE$.
The result of this process is three 
coupled equations relating the
three sets of field components
$\EoE$, $\EoB$, and $\kzE$:
\begin{widetext}
  \begin{align}
    0 &= \ddr \EoE 
    + \Frac{p^2_j}{\om+i\si}\big(\dr\Frac{\om+i\si}{p^2_j}\big) \dr \EoE + p^2_j\EoE
    +\Frac{ip^2_j}{\om+i\si}\dr \Frac{1}{rp^2_j}\kzE 
    +2i|\mbf\k|\Frac{m\om}{(\om+i\si)j(j+1)} \kzE
    \notag \\ &\quad
    -2(\k)_0 \Frac{p^2_j}{(\om+i\si)\om} \dr\EoB
    -2|\mbf\k|\Frac{p^2_jm}{(\om+i\si)j(j+1)}\EoE
    +2|\mbf\k|\Frac{m}{(\om+i\si)r} \dr\EoE
    \notag \\  &\quad
    +2i|\mbf\k|\Frac{p^2_j(j^2+2j-1){\mathcal C}_{jm}}{(\om+i\si)j(j+1)} \EoBjm{(j-1)m}
    +2i|\mbf\k|\Frac{p^2_j(j^2-2){\mathcal C}_{(j+1)m}}{(\om+i\si)j(j+1)} \EoBjm{(j+1)m} \ ,
    \displaybreak[0]
    \label{E_eq}\\
    0 &= \ddr \EoB + p^2_j \EoB 
    -2|\mbf\k|\Frac{m\om}{j(j+1)}\EoB
    +2(\k)_0 \Frac{(\om+i\si)\om}{p^2_j} \dr \EoE
    +2i(\k)_0 \Frac{\om}{p^2_j r} \kzE
    \notag \\ &\quad
    -2i|\mbf\k| \Frac{(j^2+2j-1){\mathcal C}_{jm}\om}{j(j+1)} \EoEjm{(j-1)m}
    -2i|\mbf\k| \Frac{(j^2-2){\mathcal C}_{(j+1)m}\om}{j(j+1)} \EoEjm{(j+1)m}
    -2i|\mbf\k| \Frac{(j-1){\mathcal C}_{jm}\om}{p^2_{j-1}r} \dr \EoEjm{(j-1)m}
    \notag \\ &\quad
    +2|\mbf\k| \Frac{{\mathcal C}_{jm}\om^2}{jp^2_{j-1}} \kzEjm{(j-1)m}
    +2i|\mbf\k| \Frac{(j+2){\mathcal C}_{(j+1)m}\om}{p^2_{j+1}r} \dr \EoEjm{(j+1)m}
    -2|\mbf\k| \Frac{{\mathcal C}_{(j+1)m}\om^2}{(j+1)p^2_{j+1}} \kzEjm{(j+1)m} \ ,
    \displaybreak[0]
    \label{B_eq}\\
    0 &=  \kzE  - 2i (\k)_0 \Frac{j(j+1)}{\om r} \EoB  - 2i|\mbf\k| m\EoE 
    \notag \\ &\quad
    - 2|\mbf\k| (j-1){\mathcal C}_{jm}\EoBjm{(j-1)m}
    + 2|\mbf\k| (j+2){\mathcal C}_{(j+1)m}\EoBjm{(j+1)m} \ ,
    \label{K_eq}
  \end{align}
\end{widetext}
where we define $p^2_j=\om(\om+i\si)-j(j+1)/r^2$.
Note that we could use Eq.\ \rf{K_eq}
to eliminate $\kzE$.
However, for simplicity,
we treat $\kzE$ as a dynamical
field on equal footing with $\EoE$ and $\EoB$.
The field components $\EoE$ and $\EoB$
correspond to the transverse part
of the electric field,
which vanishes at the surfaces
of a perfect conductor.
This implies 
$\kzE$ vanishes on the surfaces as well.
So we take
$\EoE$, $\EoB$, and $\kzE$
as independent fields that vanish
at the boundaries of the cavity.

\subsection{Numerical frequencies}

We calculate the resonant frequencies
that result from the modified electrodynamics
by considering discrete
radii $r_n=R+\de r (n+\half)$,
where $n$ is an integer.
Defining discrete field coefficients
at these points,
$\EoEjm{njm}$, $\EoBjm{njm}$, and $\kzEjm{njm}$,
and using discrete derivatives,
wave equations \rf{E_eq}-\rf{K_eq}
can be written in the form of an
infinite-dimensional matrix equation.
Discrete resonances correspond to frequencies
where nontrivial field configurations exist.
These can be estimated by truncating
the matrix at finite index values 
and searching for $\om$ where
the truncated matrix is singular.
These $\om$ are complex in general.
The real parts give the
resonant frequencies $\nu={\rm Re}\, \om/(2\pi)$,
while the ratios of the real and 
imaginary parts determine
the quality factors 
$Q= -{\rm Re}\, \om/{\rm Im}\, \om/2$
of the modes.

In the event that $(\k)_\ka =0$,
Eqs.\ \rf{E_eq}-\rf{K_eq} reduce to two wave equations,
one for $B$ modes and one for the Schumann $E$ modes.
However, nonzero $(\k)_\ka$ coefficients
mix the two parities,
and resonances will no longer
possess definite
$E$ or $B$ parity.
We also note that no mixing of fields with
different $m$ values occurs,
but mixing across $j$ values results from 
the vector $\mbf\k$, as expected. 
As a result,
all resonances have definite $m$ values,
and $m$ may be fixed in the calculation.

To determine the resonances,
we take 100 different $r$ values,
$n=0,\ldots 99$,
uniformly spaced between
$R$ and $1.01 R$.
This corresponds to a penetration
of about $0.001 R\simeq 6.4$ km
into the highly conductive ionosphere.
For fixed $m$,
we create a matrix including terms
corresponding to these $n$ values
and the ten lowest $j$ values
that are relevant, $j \geq |m|$.
The result is a square matrix
with dimension $100\times 10 \times 3$.
We use a row-reduction method
to determine its determinant
for different values of $\om$
and search for roots.

\section{Results}
\label{sec_results}

\begin{figure}
  \includegraphics[width=\columnwidth]{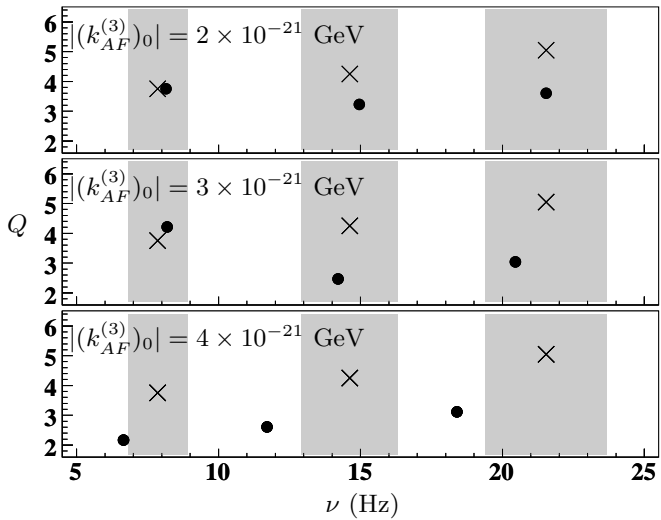}
  \caption{\label{k0fig}
    Calculated resonant frequencies
    and quality factors for three different
    values of $|(\k)_0|$.
    Circles represent the resonances.
    The Lorentz-invariant limit
    is shown with $\times$ symbols
    for comparison.
    The shaded regions indicate the
    corresponding widths at half maximum
    for the three lowest resonances
    in the Lorentz-invariant case.}
\end{figure}

\begin{figure}
  \includegraphics[width=\columnwidth]{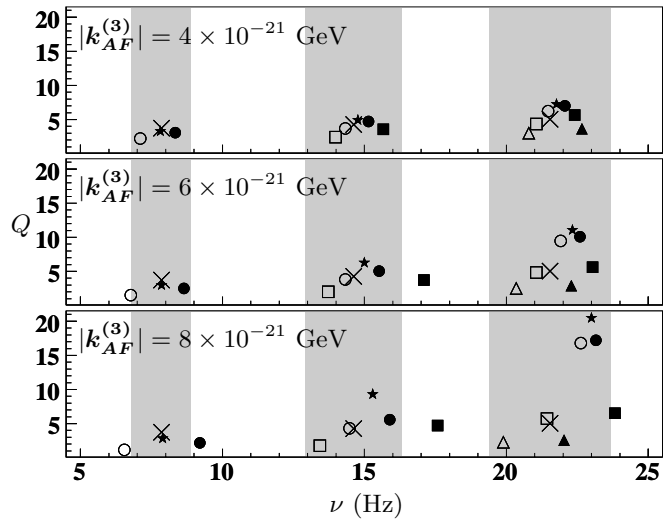}
  \caption{\label{kfig}
    Resonant frequencies and $Q$ factors for
    three values of $|\mbf\k|$.
    The $|\mbf\k|=0$ case ($\times$ symbols)
    is shown for comparison.
    The shaded regions indicate the
    corresponding widths at half maximum
    for the three lowest resonances
    in the Lorentz-invariant case.
    Each plot includes resonances
    for index values
    $m=0$ (stars),
    $m=\pm 1$ (filled/empty circles),
    $m=\pm 2$ (filled/empty squares),
    and
    $m=\pm 3$ (filled/empty triangles).}
\end{figure}

While any combination of
coefficients for Lorentz violation
is possible, for simplicity,
we next consider the effects
of $(\k)_0$ and $|\mbf \k |$ separately.
We first consider a nonzero
$(\k)_0$ coefficient.
Figure \ref{k0fig}
shows the three lowest-frequency
resonances for three values
of $|(\k)_0|$.
The three resonances shown
correspond to $j=1,2,3$
and are degenerate in $m$.
The results are independent
of the sign of $(\k)_0$ due to
the symmetry of this case.
The figure also shows the
calculated resonant frequency
and $Q$ factor for the Lorentz-invariant
limit.
These are in good agreement
with the observed resonances
of
$(\nu,Q) = (7.8\,{\rm Hz},4.0)$,
$(14.1\,{\rm Hz},4.5)$,
$(20.3\,{\rm Hz},5.0)$
\cite{schumdata}.

From the figure we see
that values of $(\k)_0$
on the order of $10^{-21}$ GeV
significantly affect
both the resonant frequencies
and $Q$ factors.
In particular,
values of $4\times 10^{-21}$ GeV
drastically alter
all three modes.
We therefore adopt a conservative limit of
$|(\k)_T| < 4\times 10^{-21}$
on the time-like part of $(\k)_\ka$
in the standard Sun-center frame
described in Ref.\ \cite{km}.
This translates to a bound of
\beq 
|\kVdjm{3}{00}| < 14\times 10^{-21}
\eeq
on the spherical coefficient
from Ref.\ \cite{km_cmb}.

Note that the above bound is
about two orders of magnitude
larger than the naive prediction.
This can be understood from
the fact that the 
$(\k)_\ka$ coefficients mix
$E$- and $B$-parity modes.
The $B$-parity resonances
have frequencies that
are at much higher frequencies.
This leads to a seesaw effect
in the Maxwell equations
that suppresses perturbations
in the resonances.
As a result,
relatively large mixing
in the wave equations must
be present for significant
changes to manifest
in the low-frequency modes.

Small changes to the conductivity
profile can lead to large changes
in the resonances,
implying that our confidence in the
bound on $(\k)_T$ is somewhat weakened by
our knowledge of the atmospheric conductivity,
which is a complicated and variable system.
Much cleaner bounds can be
placed on the pseudovector part
$\mbf\k$ since it leads to a
breakdown of the usual
$2j+1$ degeneracy among modes
with identical $j$ eigenvalues.
These bounds too are complicated by
imperfections in our conductivity model,
including the missing day-side/night-side
and polar asymmetries
that are present in the real atmosphere 
\cite{schum2}.
These conventional anisotropies
can also break the degeneracies,
but they would only add to the effects
we are bounding.
We therefore neglect them
in our analysis.
However,
these features could be significant
in more detailed studies involving
field configurations.
For example,
local fields may experience variability
that includes daily and annual fluctuations
from conventional physics.
Extracting a sidereal dependence
caused by the rotation of the Earth
with respect to the fixed $\mbf\k$
vector might be possible
but is beyond the scope of this work.

Figure \ref{kfig} shows the resonances
for three different values of $|\mbf\k|$.
We notice shifts in the frequencies and
$Q$ factors as well as the expected
$2j+1$ splitting of the resonances.
In particular, values of 
$|\mbf\k| = 8\times 10^{-21}$ GeV
lead to new resonances separated
by frequency intervals comparable
to the resonance widths.
These multiple resonances would
be evident in the data if they existed.
Therefore, we take 
$|\mbf\k| < 8\times 10^{-21}$ GeV
as a conservative bound.
The relation to the spherical coefficients
is given by $|\mbf{\kafd{3}}|=\frac{1}{\sqrt{4\pi}}
\big(6|\kVdjm{3}{11}|^2+3|\kVdjm{3}{10}|^2\big)^{1/2}$
\cite{km_astro}.
So our bound leads to two constraints on
spherical coefficients for Lorentz violation of
\begin{align}
  |\kVdjm{3}{11}| &< 12 \times 10^{-21} \mbox{ GeV} , \notag\\
  |\kVdjm{3}{10}| &< 16 \times 10^{-21} \mbox{ GeV} .
\end{align}
These completely bound the vector ($j=1$)
dimension-three Lorentz-violating operators.
Again,
these constraints are less stringent than
the naive estimate.

\section{Discussion}

In this work,
we used Schumann resonances
in the Earth-ionosphere cavity to
place bounds on the
order of $10^{-20}$ GeV
on $CPT$-odd $j=1$ coefficients
of the minimal SME.
Similar bounds are places on $j=0$
scalar coefficients assuming
the actual conductivity profile of
the atmosphere is not significantly
different from our model profile.
These bounds constitute the
first terrestrial bounds
on dimension-three Lorentz-violating
operators.

While not as sensitive as
astrophysical searches
for vacuum birefringence,
the techniques used in this work
test Lorentz invariance in
our local neighborhood,
giving a bound on coefficients
over solar-system length scales.
In contrast,
astrophysical tests probe
Lorentz violation over cosmological
scales and may be obfuscated by
spacetime variations 
or domains in the
Lorentz-violating backgrounds.
So local tests play
an important role in our
search for new physics.

Laboratory-based experiments
may also provide local tests
of Lorentz invariance.
Current cavity experiments
utilize high $Q$ factors that
allow for sensitivities to 
$\de\nu/\nu$ on the order
of parts in $10^{15}$ or better
\cite{cavities}.
This suggests improved bounds
may be possible in laboratory
experiments.
A rough estimate yields sensitives to
$(\k)_\ka$ from about
$10^{-25}$ GeV
in optical cavities to around
$10^{-30}$ GeV
in lower-frequency microwave cavities.

Future studies involving Schumann
resonances may be able to
improve on the above bounds.
Precise tracking of the resonances
may allow for sidereal searches
that would indicate rotation violations
from $\mbf\k$.
Also, boost violations from $(\k)_\ka$ coefficients
can lead to annual dependences that may
be discernible.
Regardless,
the constraints obtained here demonstrate
the potential of resonator experiments
as tests of dimension-three Lorentz violations
in the atmosphere and in the laboratory.

\appendix 

\section{Spin weight}
\label{sec_sw}

The problem addressed in this work
is most naturally solved in
spherical coordinates.
Here we use a helicity-based system
and covariant-angular-momentum operators.
In this appendix,
we summarize some the key identities
used in the calculation of the modified
wave equations.
A fuller discussion of these methods
will appear elsewhere
\cite{km2}.
The technique is based on a decomposition
into total angular momentum $\mbf J$
and helicity.
A type of tensor spherical harmonics
called spin-weighted harmonics $\syjm{s}{jm}$
\cite{sYjm}
provide orthonormal sets of eigenfunctions
of these operators.
The index $s$ labels the spin weight,
which up to a sign is equivalent to helicity.

The method starts by defining 
helicity basis vectors,
$\mbf{\hat e}_\pm ={\mbf{\hat e}}^\mp
= \Frac{1}{\sqrt{2}}(\mbf{\hat e}_\th \pm i \mbf{\hat e}_\ph)$,
$\mbf{\hat e}_r=\mbf{\hat e}^r$,
where $\mbf{\hat e}_r = \mbf x/|\mbf x|$
is the radial unit vector, and
$\mbf{\hat e}_\th$ and
$\mbf{\hat e}_\ph$ are unit vectors
associated with the usual coordinate
angles $\th$ and $\ph$.
The helicity operator 
$J_r=\mbf{\hat e}_r\cdot \mbf J$
generates local rotations
about the radial direction.
In the helicity basis,
components of 3-dimensional tensors
have definite spin weight,
which can be determined
by counting the number of
$+$ and $-$ index values.
For example,
consider a tensor component 
$T_{++}=T^{--}=\mbf{\hat e}_+\cdot T \cdot\mbf{\hat e}_+$.
It is a spin-weight-2 function
and can be expanded into the complete set
of spin-weight-2 spherical harmonics $\syjm{2}{jm}$.
Another example is the tensor component 
$T_{-r}=\mbf{\hat e}_-\cdot T \cdot\mbf{\hat e}_r$,
which has a spin weight of $-1$.
In the present context,
the electric and magnetic fields
have spin-weight-0 components
$E_r = \mbf{\hat e}_r\cdot \mbf E$,
$B_r = \mbf{\hat e}_r\cdot \mbf B$
and components 
$E_\pm = \mbf{\hat e}_\pm\cdot \mbf E$,
$B_\pm = \mbf{\hat e}_\pm\cdot \mbf B$,
which have a spin weight of $\pm 1$.

In general,
harmonics of a given weight satisfy
completion and orthogonality relations,
\begin{align}
  \sum_{jm} \syjm{s}{jm}^*(\Om)\, \syjm{s}{jm}(\Om')
  &= \de(\Om-\Om')\ , \\
  \int  \syjm{s}{jm}^*(\Om)\, \syjm{s}{j'm'}(\Om)\, d\Om
  &= \de_{jj'}\de_{mm'} \ .
\end{align}
More generally, they obey
\begin{align}
  \syjm{s_1}{j_1m_1}\syjm{s_2}{j_2m_2}
  =\ & \sum_{s_3j_3m_3} \sqrt{\Frac{(2j_1+1)(2j_2+1)}{4\pi(2j_3+1)}}
  \notag\\
  &\times \cg{j_1j_2(-s_1)(-s_2)}{j_3 (-s_3)}
  \notag\\
  &\times \cg{j_1j_2m_1m_2}{j_3 m_3}
  \ \syjm{s_3}{j_3m_3}\ ,
  \label{orth}
\end{align}
where $\cg{j_1j_2m_1m_2}{j_3 m_3}$
are Clebsch-Gordan coefficients.
Note that orthonormality and completion
do not extend across harmonics of different
spin weight.

The spin-weighted harmonics
can be generated for the familiar
spin-weight-zero harmonics 
$\syjm{0}{jm}=Y_{jm}$
using covariant-angular-momentum operators
$J_a = -i \ep_{abc} x^b \nabla^c + S_a$,
where $\nabla^a$ represent covariant
derivatives and $S_a$ are covariant
spin operators.
The spin operators act on tensors
in a manner similar to a connection.
For example,
operating on a tensor
${T^b}_c$, we get
$S_a {T^b}_c= S^b_{ad} {T^d}_c - S^d_{ac} {T^b}_d$,
where $S^c_{ab}=i{\ve^c}_{ab}$.
The totally antisymmetric tensor has
nonzero components
$\ve_{+r-} = -\ve^{+r-} = i$
in the helicity basis.
In general,
the components $J_\pm=\mbf{\hat e}_\pm \cdot \mbf J$
raise/lower the spin weight
of a function.
Specifically acting on the spherical harmonics,
we get the ladder relations
\beq
J_\pm\ \syjm{s}{jm}= 
-\sqrt{\half\big(j(j+1)-s(s\pm 1)\big)}
\syjm{s\pm1}{jm} \ .
\eeq
Successive operations
generate arbitrary $\syjm{s}{jm}$
in terms of $\syjm{0}{jm}$.

Finally,
there is a useful relationship between
the covariant derivatives
and the angular-momentum operators
in the helicity basis:
\begin{align}
  \nabla_r &= \prt/\prt r \ ,   \notag \\
  \nabla_\pm &= \pm(J_\pm - S_\pm)/r \ . \label{J-ident2}
\end{align}
These identities help in the
reduction of differential tensor
equations into radial
and angular parts.
For example,
the divergence of vector $V^a$ becomes
$\nabla_a V^a  
= \big(\Frac{\prt \phantom{r}}{\prt r} +\Frac{2}{r}\big)V_r 
+ \Frac{1}{r}(J_+V_- - J_-V_+)$,
using the properties of the spin operator.

\end{document}